\documentclass[epj]{svjour}
\usepackage{graphics}
\begin{document}
\title{GARFIELD + RCo Digital Upgrade: a Modern Set-up for Mass and Charge Identification of Heavy Ion Reaction Products}
\author{
M. Bruno\inst{1} \and
F. Gramegna\inst{2} \and
T. Marchi\inst{2,3} \and
L. Morelli\inst{1} \and
G. Pasquali\inst{4,5} \and
G. Casini\inst{4} \and
U.Abbondanno\inst{6} \and
G. Baiocco\inst{1}\thanks{\emph{Present address: INFN, Sezione di Pavia, Italy}} \and
L. Bardelli\inst{4,5} \and
S. Barlini\inst{4,5} \and
M. Bini\inst{4,5} \and
S. Carboni\inst{4,5} \and
M. Cinausero\inst{2} \and
M. D'Agostino\inst{1} \and
M. Degerlier\inst{2}\thanks{\emph{Present address: Physics Department, Nevsehir University, Turkey}}\and
V. L. Kravchuk\inst{2}\thanks{\emph{Present address: National Research Centre "Kurchatov Institute", Moscow, Russia
}} \and
E. Geraci\inst{1}\thanks{\emph{Present address: Dipartimento di Fisica ed Astronomia and INFN, Catania, Italy}}  \and
P. F. Mastinu\inst{2} \and
A. Ordine\inst{7} \and
S. Piantelli\inst{4}\and
G. Poggi\inst{4,5} \and
A. Moroni\inst{8}\thanks{\emph{Present address: Malgesso (Varese), Italy}}
}                     
%
%
\institute{Istituto Nazionale di Fisica Nucleare, Sezione di Bologna and \\ Dipartimento di Fisica ed Astronomia dell'Universit\`a, Bologna, Italy \and Istituto Nazionale di Fisica Nucleare, Laboratori Nazionali di Legnaro, Italy \and
Dipartimento di Fisica ed Astronomia dell'Universit\`a, Padova, Italy \and
Istituto Nazionale di Fisica Nucleare, Sezione di Firenze, Italy \and
Dipartimento di Fisica ed Astronomia dell'Universit\`a, Firenze, Italy \and
Istituto Nazionale di Fisica Nucleare, Sezione di Trieste, Italy \and
Istituto Nazionale di Fisica Nucleare, Sezione di Napoli, Italy \and
Istituto Nazionale di Fisica Nucleare, Sezione di Milano, Italy}
\date{Received: date / Revised version: date}
%
\abstract{
An upgraded GARFIELD + Ring Counter (RCo) apparatus is presented with improved performances as far as electronics and detectors are concerned. On one side fast sampling digital read out has been extended to all detectors, allowing for an important simplification of the signal processing chain together with an enriched extracted information. On the other side a relevant improvement has been made in the forward part of the setup (RCo): an increased granularity of the CsI(Tl) crystals and a higher homogeneity in the silicon detector resistivity. The renewed performances of the GARFIELD + RCo array make it suitable for nuclear reaction measurements both with stable and with Radioactive Ion Beams (RIB), like the ones foreseen for the SPES facility, where the Physics of Isospin can be studied.
} 
\authorrunning {M. Bruno et al.}
\titlerunning {GARFIELD + RCo Digital Upgrade}
\maketitle
\section {Introduction} \vspace {0.1cm}

The international nuclear physics community is more and more interested in the development of new facilities based on accelerated radioactive ion beams~\cite{nupecc}. Since many years several laboratories worldwide are active in this field. Limiting at the European level one can mention: the FAIR project~\cite{fair} at GSI in Germany, Spiral II~\cite{sp2} at GANIL in France, HIE-ISOLDE~\cite{hie} at CERN, FRIBS~\cite{fribs} and EXCYT~\cite{excyt} at INFN Laboratori del Sud (LNS) in Catania and SPES~\cite{spes} at INFN Laboratori Nazionali di Legnaro (LNL) in Italy. Together with the construction of these facilities and in view of the forthcoming EURISOL project~\cite{eurisol}, new instrumentation has to be developed and the existing detectors have to be upgraded. In this framework, focusing on the charged particle detectors and their use at moderate energy in RIB facilities, it is important to achieve good isotopic separation in a wide range of fragment masses together with low energy identification thresholds.

\begin{figure*}[htb]
\centering
\resizebox{0.80\textwidth}{!}{%
  \includegraphics{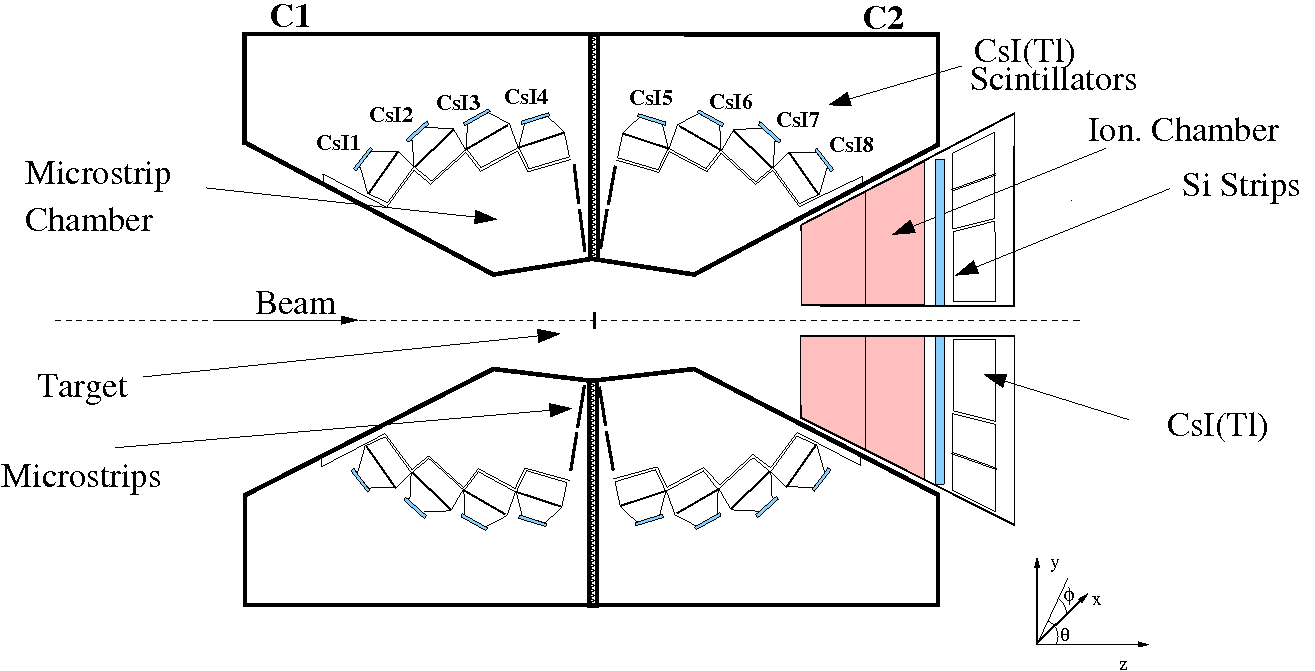}}
 \caption {\it (Color online) Schematic view of the two GARFIELD + RCo array.}\label{tot}
 \centering
\end{figure*}

 Several apparatuses exist in different laboratories with this aim. For example CHIMERA~\cite{chimera} at LNS is a multi-detector based on 1192 silicon - CsI(Tl) telescopes exploiting Time of Flight (ToF), $\Delta E-E$ and scintillator fast-slow technique to detect and identify charged particles. At GANIL the INDRA apparatus~\cite{indra} features a Ionization Chamber (IC) stage in front of silicon detector and CsI(Tl) scintillators to reduce the energy threshold for charge identification. It detects the products emitted in nuclear reactions both with stable and unstable (Spiral) beams. Both these apparatuses cover the whole solid angle and presently rely mainly on analog electronics. They obtain mass and charge identification for light fragments (Z up to 3$\div$4) through fast-slow Pulse Shape Analysis (PSA) in CsI(Tl) while silicon PSA has been obtained in CHIME\-RA via specific modules allowing the double CFD method.
Apparatuses equipped with digital electronics exist in US, like NIMROD~\cite{nimrod} in Texas A\&M University Cyclotron and HIRA~\cite{hira} at the NSCL of MSU, which exploit several available particle identification techniques including PSA in silicon detectors~\cite{psa}. Finally one has to mention the FAZIA demonstrator~\cite{fazia_web}; this array of 192 three-stage Si-Si-CsI modules is currently under construction. It presents the best performances reachable with this kind of detectors, thanks to specific detector solutions and a new designed digital electronics. FAZIA is a versatile system able to be coupled with other large arrays.

In order to meet the more demanding performances in experiments with the future RIB facility at LNL, the GARFIELD + Ring Coun\-ter (RCo)~\cite{garf,rco} charged particle and fragment detection array has been upgraded in order to obtain both light charged particle (LCP, Z~$\le$~2) isotopic separation in all covered range (5$^o$ -150$^o$) and a better position determination (granularity) at the forward angles where the kinematics concentrates most of the reaction cross section. Moreover particular care has been devoted to increase the fragment (Z~$\ge$~3) identification capabilities. This upgrade allows for a more complete event reconstruction and therefore a better characterization of the different competing reaction mechanisms.

The GARFIELD array consists of two independent gas drift chambers especially designed to identify intermediate mass fragments and light charged particles. They exploit different techniques based on  micro-strip gas chambers~\cite{glass} and CsI(Tl) scintillators~\cite{wag,mor}.
The forward angular range is covered by the RCo, a detector based on a three-layer telescope configuration: Ionization Chamber (IC), Silicon Strip Detectors and CsI(Tl) scintillators.
In Fig.~\ref{tot} a sketch of the complete GARFIELD + RCo apparatus is shown. In this configuration a geometrical efficiency of $\sim 80\%$ is achieved.

According to the needs of different experiments,
 different combinations of GARFIELD, RCo and ancillary detectors have been used. In particular a time of flight system composed by large area
position sensitive parallel plate avalanche counters
(PSPPAC) has been used instead of the RCo to detect evaporation residues and/or fission fragments in
fusion/fission nuclear reactions. This allowed to perform experiments aimed at a detailed study of light particle pre-equilibrium emission~\cite{krav}.
Further experiments have been performed by using
a set of phoswich detectors made of three different
scintillation layers (two plastics and one CsI(Tl) crystal)
read by a unique photomultiplier, formerly used in the FIASCO
experiment~\cite{phos}.
For this kind of experiments a system of BaF$_2$ scintillators (HECTOR apparatus~\cite{hect})
was used in place of the backward drift chamber: this
coupling was considered in order to detect high energy gammas in
coincidence with evaporation residues and with light charged
particles. With this set up we have studied the Giant Dipole
Resonance~\cite{giant}, the isospin mixing~\cite{isomix}, the
Jacobi shape transition~\cite{jac} and the dynamical dipole
emission~\cite{dip}. Recently the first version of the GARFIELD + RCo set up has been used to study
odd-even effects in the isotopic fragment distribution (staggering) up to the multi-fragmentation threshold~\cite{stag}. The upgraded set up has been recently exploited in low energy light nuclei reactions to get information on the statistical and non-statistical behavior of light particle emission mechanism~\cite{cc}.
In the next future
the FAZIA~\cite{fazia_web} telescopes could be coupled to GARFIELD,
to perform nuclear dynamics and thermodynamics studies in the
perspective of the RIB of SPES.

In sections~\ref{garf} and \ref{rco} the main features of the setup will be recalled and the major upgrades, which are hereafter listed, will be described in detail:
\begin{enumerate}
\item design and implementation of new digitizing Front End Electronics (FEE) for all detector channels~\cite{dig1}, along the same lines of what has been done for new generation apparatuses such for instance HIRA~\cite{hira} for charged particles or AGATA for $\gamma$ rays~\cite{agata};
\item replacement of the old RCo Silicon detectors with neu\-tron-transmutation-doped ones (nTD)~\cite{can};
\item increase of the granularity of the RCo CsI(Tl) scintillators.
\end {enumerate}

\section{GARFIELD}\label{garf}

\begin{figure}[htb]
\centering
\resizebox{0.38\textwidth}{!}{%
  \includegraphics{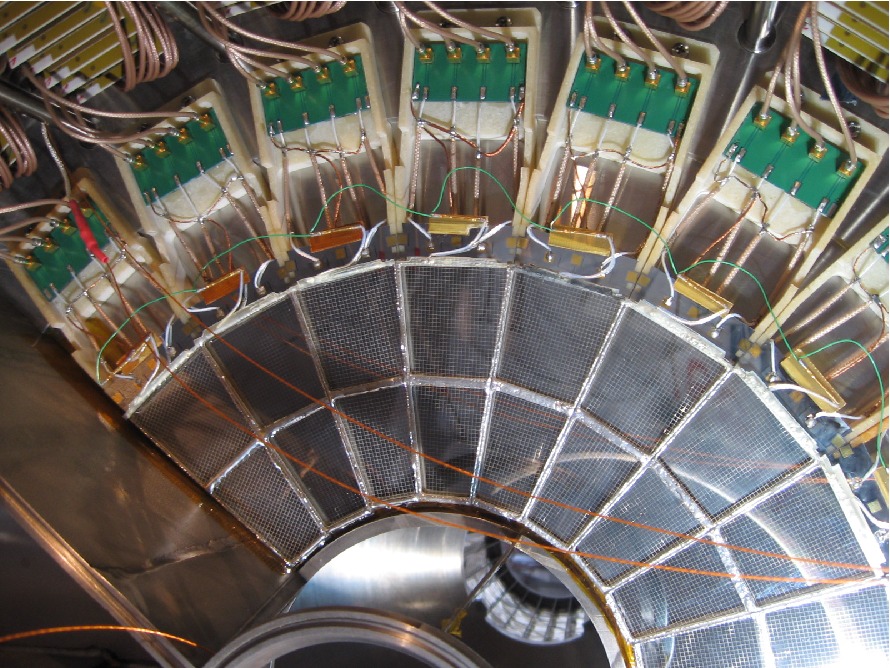}}
\caption
{\it (Color online) Picture of the micro-strip pads mounted in one of the GARFIELD chambers. Cabling and pre-amplifiers are visible in the upper part of the figure.}\label{micro1}
\centering
\end{figure}

\begin{figure}[htb]
\centering
\resizebox{0.5\textwidth}{!}{%
  \includegraphics{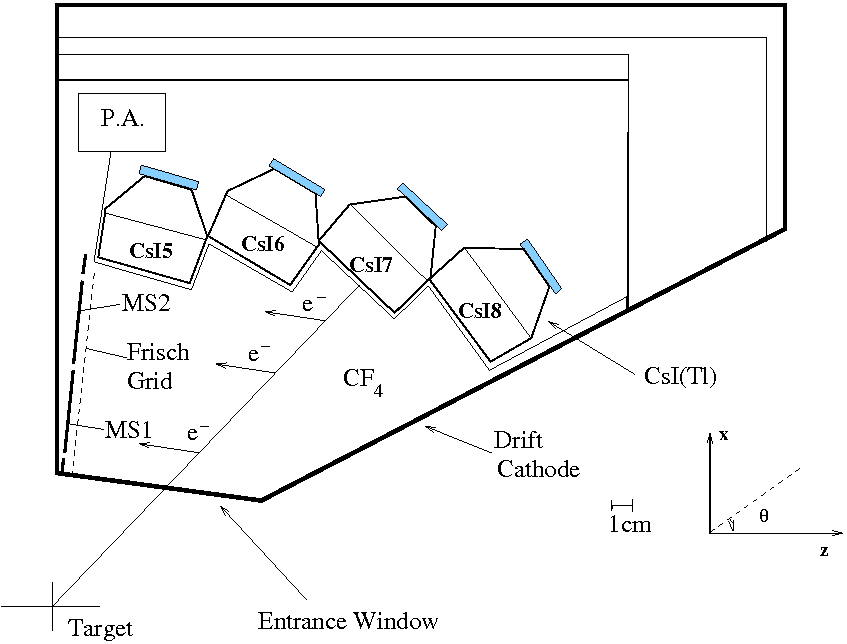}}
\caption
{\it (Color online) Schematic view of one of the GARFIELD sector. The electrons produced along the ion track drift towards the micro-strip pads allowing the measurement of the energy loss and of the drift time. The residual energy is measured by the CsI(Tl) scintillators. P.A. represents the motherboard carrying 8 preamplifiers.}\label{driftfig}
\centering
\end{figure}

The GARFIELD apparatus consists of two large volume gas detectors employing micro-strip gas chambers (MSGC) as $\Delta E$ stage, complemented by CsI(Tl) scintillators as residual energy detectors. This allows to achieve low identification thresholds (0.8$\div$1~MeV/A) using the $\Delta E - E$ technique. The two drift chambers have cylindrical symmetry and are placed back to back with respect to the target. The first chamber (C1 in Fig.~\ref{tot}) covers the angular region from $\theta= 95 ^\circ$ to $\theta= 150 ^\circ$ and has a side gap of $\Delta \phi \simeq 45^\circ$ which allows the placement of different ancillary detectors. The second one covers the angular region from $\theta= 30 ^\circ$ to $\theta= 85 ^\circ$ and the complete azimuthal angle.  The two gas volumes are respectively divided in 21 and 24 azimuthal sectors and are filled with CF$_4$ at a typical pressure of 50 mbar. Each sector is equipped with one trapezoidal micro-strip pad~\cite{glass} divided in 4 different collecting zones providing the $\Delta E$ signals (see section~\ref{mcgs}). The ensemble of all 21 (24) trapezoidal pads forms a ''ring`` around the target (see Fig.~\ref{micro1}). The pads are positioned almost perpendicularly ($82^\circ$) to the  beam axis (see Fig.~\ref{driftfig}). A total of 84 (96) micro-strip signals are read out in the first and the second chamber respectively.\\
As shown in Fig.~\ref{driftfig}, each sector is completed by four CsI(Tl) crystals with photodiode readout which provide the residual energy signals ($E$). These detectors also define the boundaries of the sensitive gas volume and are positioned in a radial configuration with respect to the interaction point (see section~\ref{csigarf}).\\
In each sector one motherboard carrying 8 pre-amplifiers is mounted inside the gas volume, close to the detectors in order to minimize their input capacitance. The pre-amplifiers have been designed to have a large dynamic range (up to 10 V) and low noise, less than 130 ENC (Equivalent Noise Charge), together with low power consumption (about 200 mW each)~\cite{pre}. Their nominal gain is 45 mV/MeV silicon equivalent. A  water recirculating cooling system is used to dissipate the pre-amplifiers generated heat.\\

The original version of the array was equipped with an analog electronic chain mainly composed by the CAEN N568B shaping amplifiers, SILENA FAIR 9418/6-V32 peak sensing ADCs, SILENA FAIR 9418/6-T32 TDCs and C208 CAEN Camac CFDs. ADCs and TDCs were part of a VME acquisition system, based on FAIR protocol~\cite{fairna}. Also VME delays based on the FAIR protocol were part of the electronic chain. The complete analog chain has been replaced by new digitizing Front End Electronics (FEE) ad-hoc developed by the collaboration, also compatible with the FAIR framework.

The core of the new FEE is a single channel board based on a fast sampling ADC followed by a Digital Signal Processor (DSP). The main features of the 125 MHz - 12 bit digitizing cards are described in~\cite{dig1}. The DSP is used to perform on-line Pulse Shape Analysis (PSA)~\cite{ben} and extract the relevant parameters from the sampled signal: energy, timing and pulse-shape information are obtained using the algorithms described in~\cite{dig1,dig2,dig3} and later summarized (see section~\ref{algo}).

\subsection{Gas micro-strip detectors}\label{mcgs}

\begin{figure}[htb]
\centering
\resizebox{0.5\textwidth}{!}{%
  \includegraphics{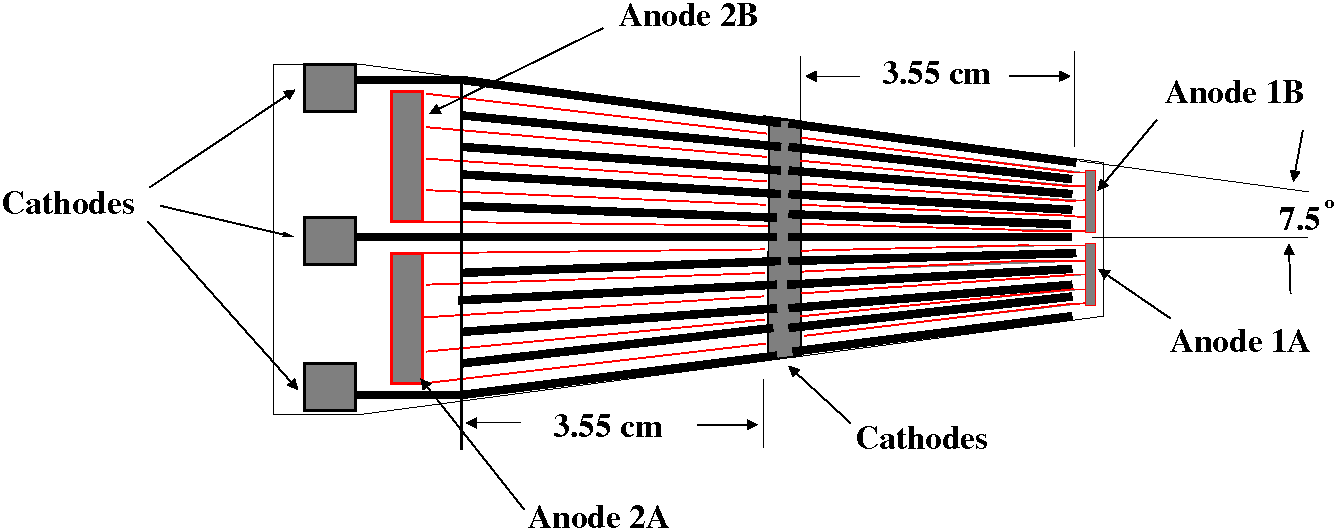}}  \caption
{\it (Color online) A sketched micro-strip pad corresponding to one of the GARFIELD sectors.}\label{micro}
\centering
\end{figure}

Each GARFIELD micro-strip pad has a keystone geometry and it is divided in four collecting zones (1A, 1B, 2A, 2B, as sketched in Fig.~\ref{micro}). The pads are composed of very small anode/cathode alternated metallic electrodes (distance = 50 $\mu m$) that are deposited through photo-lithography technique on glass. The anodes are $10~\mu m$ large, electrically connected into the four collecting areas and biased at 430 V. The cathodes have a trapezoidal shape (width from 85~$\mu m$ to 140~$\mu m$ in the 1A and 1B zones and from 85~$\mu m$ to 190~$\mu m$ in the 2A and 2B ones) and are all grounded.

The MSGC operation is based on the avalanche multiplication process in gas, which is used to amplify the signals due to primary ionization electrons. A proper electric field forces the free electrons formed along the track of the impinging particle to drift toward the anode. The anodic plane is screened towards the drift region by a Frisch grid positioned 3 mm apart: in this way the primary electrons start feeling the very strong MSGC electric field ($\simeq 10^7$ V/m) only after crossing the grid, thus producing the electron avalanche. Using CF$_4$ gas at 50 mbar, typical gain factors of 30-50 are achieved: this is one of the main advantage of using MSGC with respect to normal ionization chambers. CF$_4$ gas is used due to its high density (0.19 mg/cm$^3$ at a pressure of 50 mbar and at a temperature 20 $^\circ$C) and high electron drift velocity (10 cm/$\mu$s at 1 V/cm/mbar). An automatic recirculating and filtering system makes the gas continuously flowing in the chamber to minimize the Oxygen contamination.

The obtained signals carry the information on the energy loss ($\Delta E$) and on the polar angle ($\theta$) of the impinging ion~\cite{glass,onori}. Every MSGC collecting zone allows for an angular resolution of $\Delta \phi = 7.5^\circ$ and $\Delta \theta = 1^\circ\div3^\circ$~\cite{garf}, the latter exploiting the measured drift time of the electrons. The energy resolution of the chambers is about 6\%~\cite{garf} for elastically scattered $^{32}$S at 200 MeV leaving about 10 MeV in the gas.

\subsection{CsI(Tl) scintillators}\label{csigarf}

As shown in Fig. \ref{driftfig} each GARFIELD sector contains 4 CsI(Tl) scintillators, the shape of which has been designed in order to optimize the geometrical efficiency. A total of 96 (84) crystals are present in the forward (backward) chamber. Each detector covers $\Delta \phi = 15^\circ$ and $\Delta \theta \simeq 15^\circ $ ($<\theta_{forw}>$ = 35$^\circ$, 47$^\circ$, 60$^\circ$, 75$^\circ$;
$<\theta_{back}>$ = 105$^\circ$, 120$^\circ$, 133$^\circ$, 145$^\circ$). The crystal thickness is 4 cm in order to stop the charged products expected in the energy range of interest. The Thallium doping is about 1200 ppm and contributes, together with the chosen diffusive wrapping, to maximize the light collection~\cite{garf,ton}. The CsI(Tl) crystals are optically coupled to Hamamatsu S3204-05 photodiodes closely connected to the pre-amplifiers. The energy resolution is about 3\% for 5 MeV $\alpha$ particles and 2$\div$3\% for 8~MeV/u Li and C elastically scattered beams on Au target.

\subsection{Digital FEE Algorithms}\label{algo}

As already stated, the major upgrade of the GARFIELD array is related to the digital readout electronic chain~\cite{dig1}. Each pre-amplifier output signal is now directly fed into a fast sampling ADC and on-line processed by a DSP. The algorithm implemented on the signal processor extracts time, energy and particle identification information from the sampled signal. According to the different detector type specific calculations are performed.

The energy information is derived from the maximum amplitude of a semi-Gaussian shaper with a selectable value of peaking time (usually 4-5 $\mu$s). This shaper has been designed to faithfully reproduce the response of the analog shaping amplifier previously employed. The shaper is an Infinite Impulse Response filter, implemented using a recursive algorithm. In order to reduce truncation error propagation on our 16-bit DSP, the algorithm exploits double precision calculations (32-bit) and to reduce calculation time the input signal is decimated 16-fold before filtering, thus increasing its sampling period from 8 to 128~ns. Since the ballistic deficit of the digital version, as a function of the signal rise-time, is the same of the analog shaper, the calibration points obtained in the past using various ion-energy combinations can still be used.

Since all the digitizing channels are triggered by the global trigger of the experiment, a zero suppression step is needed to reduce the data flow to the acquisition. This is done channel by channel inside the DSPs, where the data output is validated only if the shaper maximum amplitude is above a selected threshold. The global trigger is composed by the sum of simple (singles) or composite (coincidences) trigger signals: generally a downscaled single logic signal (i.e. the reduced logic OR of all the GARFIELD CsI(Tl) detectors and/or the logic OR of all the RCo Si detectors) plus more structured ones are used. The more complex triggers are related to all possible coincidences among different parts of the apparatus, which can be from time to time needed depending on the specific experiment [i.e (OR Garfield) \& (OR RCo)].

Timing information from the sampled signals are obtained using a digital CFD algorithm based on cubic interpolation~\cite{dig4}. This procedure calculates a digital CFD time mark (dCFD) corresponding to a specific percentage (typically 25\%) of the signal maximum amplitude. This amplitude is evaluated by the DSP as the average of the absolute maximum and the following 15 samples of the digitized signal in order to reduce the influence of electronic noise fluctuations. The 8~ns digitizer sampling period is known within $1/10^8$ and it is stable within a few units in $10^5$ so that the conversion from samples to nanoseconds is straightforward and, therefore, no time calibration is needed. Since the on-board DSP features fixed-point arithmetics and a relatively time-consuming division algorithm, the cubic interpolation, needed to extract the time mark, is best performed off-line. The DSP just calculates the dCFD threshold and it identifies the four samples needed for interpolation (two before and two after the threshold crossing) which are then sent to the data acquisition. This allows to reduce the processing dead time and to achieve a better precision in the time mark computation. The time mark is in general used to avoid contamination of different beam bunches. In the following the specific algorithms used for the different detectors will be described.

\subsubsection{Drift chamber signal analysis} \label{algdrift}

For the MSGC detectors both energy and drift-time information are needed: the energy is evaluated as previously described, while the electron drift time in the gas, which corresponds to the polar angle information, is obtained as the time difference between two dCFD time marks. They are calculated respectively on the CsI(Tl) and on the microstrip signals.

\begin{figure}[htb]
\centering
\resizebox{0.5\textwidth}{!}{%
  \includegraphics{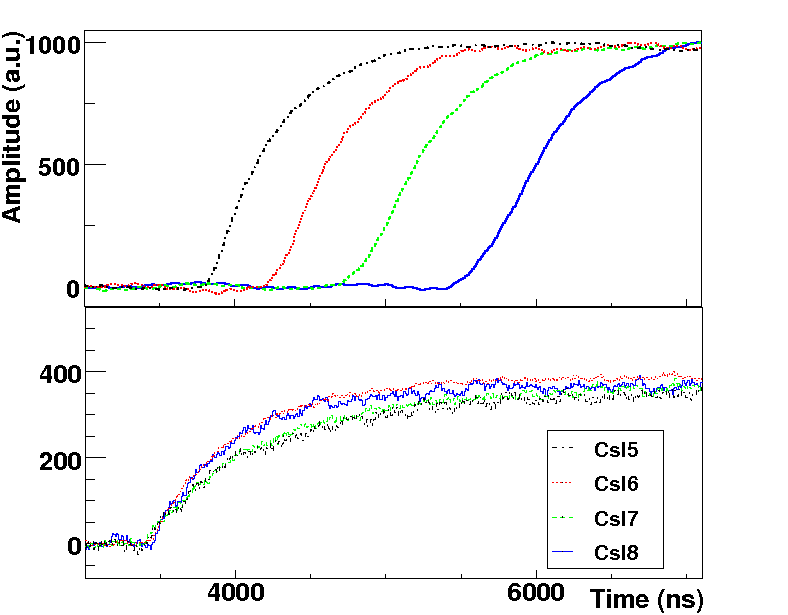}}
\caption {
{\it (Color online) GARFIELD microstrip signals and associated (coincident) CsI(Tl) signal. Bottom panel: CsI 5$\div$8 signals acquired in coincidence with microstrip signals. Top panel: Signals of a GARFIELD microstrip pad obtained in coincidence with CsI 5$\div$8 respectively (from left to right).  As expected the delay of the microstrip signal with respect to the one of the associated CsI(Tl) increases as a function of the polar angle with respect to the beam~(larger drift time).}}\label{shapes}
\centering
\end{figure}

Fig.~\ref{shapes} shows signal shapes from ions hitting a given Sector.
The tracks produce CsI(Tl)-microstrip coincidences, represented with four colors,
one for each of the four polar regions corresponding to
the CsI(Tl) crystals (from 5 to 8). Each event is triggered by one
of the CsI(Tl), thus the CsI(Tl) waveforms (bottom panel) start
at about the same time mark (there is an almost fixed delay
between the global trigger and the CsI(Tl) trigger).
However, signal pairs have been realigned in time in such a way that the 12.5\% threshold of the CsI(Tl) falls exactly at 3500ns.
Since CsI(Tl) scintillators are placed at different polar angles, the electron drift times associated to the different tracks (microstrip signal) change and in particular grow from CsI5 to CsI8 (and from CsI4 to CsI1 respectively for the C1 chamber), as one can observe in the top panel in  Fig.~\ref{shapes}.

\begin{figure}[htb]
\centering
\resizebox{0.5\textwidth}{!}{
  \includegraphics{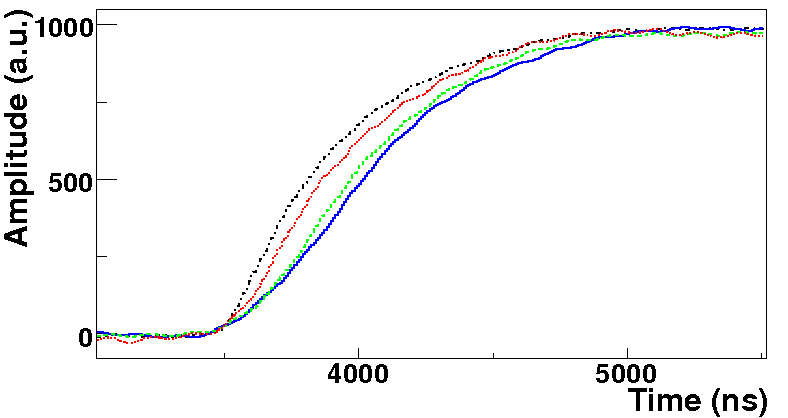}}
\caption
{\it (Color online) Average micro-strip signals associated to different CsI(Tl): signals are time-aligned in order to show their different rise-times.}\label{shapes2}
\centering
\end{figure}

Also the signal rise-time is  sensitive to the track polar angle, as shown in Fig.~\ref{shapes2}, where the signals recorded from the four collecting zones have been time aligned for a better comparison. The rise-time is evaluated as the difference between two dCFD time marks calculated on the same signal, one at 12.5\% and the other at 62.5\% of the maximum amplitude. The dependence of the rise-time on the polar angle is somewhat milder than the one of the drift time as it will be shown in Sect.~\ref{psamic}.

\rm

\subsubsection{Algorithms for the CsI(Tl)} \label{algcsig}

Time, energy and particle identification information are all needed from the CsI(Tl) detectors.
The time and energy information are obtained as described in the general section.
A further information is related to the particle identification procedure: to identify particles stopped in the CsI(Tl) crystal the well known technique of shaping the charge signal with two different time constants is used~\cite{mor}. Pulse shape identification of light charged particles is obtained correlating the amplitudes
of two semi-Gaussian shapers with a peaking time of 2.5~$\mu$s and 5~$\mu$s respectively (see Sect.~\ref{psacsgar}).

 Before the upgrade, pulse shape identification in CsI(Tl) detectors was not available, since it would have required twice the number of shaper amplifiers and ADC's. Timing information was available via standard CFD and TCD modules, which were removed after the upgrade. The use of digital electronics has thus extended the capabilities of the GARFIELD apparatus also allowing to simplify the front end electronics.

\rm
\subsection{Upgraded GARFIELD performances} \label{perfgar}

\subsubsection{$\Delta$E - E identification} \label{deegarf}

\begin{figure}[!h]
\centering
\resizebox{0.45\textwidth}{!}{%
  \includegraphics{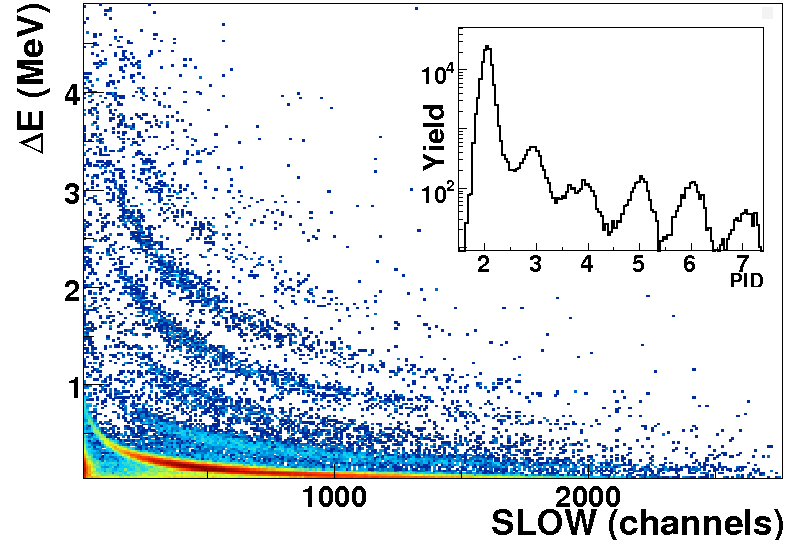}
}
\caption{\it (Color online) {$\Delta$E - E plot for the energy loss in the gas {\it vs.} the residual energy (light output) in the CsI(Tl) at $\theta$~=~47$^o$. In the inset the Particle IDentification parameter (PID), as obtained
from a linearization of the correlation, is shown. The reaction is $^{16}$O+$^{65}$Cu at 16~MeV/u beam energy.
The $\Delta$E information is obtained from one of the outer microstrips. The residual energy
is measured by CsI7 of the same sector.}}\label{deegar}
\centering
\end{figure}

The microstrip electrodes and the CsI(Tl) scintillators give a total number of 360 three stage telescopes.
The $\Delta$E-E correlation, where $\Delta$E is the energy lost in one or both sections of the drift chamber and E is the residual energy released by the incoming particle in the CsI(Tl), is shown as an example in Fig.~\ref{deegar} for the reaction $^{16}$O+$^{65}$Cu at 16~MeV/u beam energy. LCP and fragments shown in the figure are produced with charge from Z~=~1 to Z~$\approx$~8. For Z = 2 the maximum energy is around 110 MeV, whereas for Z~=~6 is around 140 MeV. Using a linearization algorithm a Particle IDentification parameter (PID) can be obtained.

The charge (mass) ion identification procedure adopted so far~\cite{mor,len,mastinu,alder,carboni_fazia} is a two-step process:
\begin{enumerate}
\item For each employed detector, in a bidimensional scatter
plot ($\Delta$E-E or fast-slow) several points are sampled
on the ridges of well defined isotopes.
\item The set of points for a given
ridge are fitted one by one using an analytical function (if it exists)~\cite{mor,len,mastinu} or via polynomial functions
(spline)~\cite{alder,carboni_fazia}.\\
\end{enumerate}
Event by event, isotopes are identified in charge (mass),
by minimizing the distance of the measured signals with
respect to the identification function.
This distance provides information on the dispersion around the most probable value, which corresponds to the charge (mass) value.

As an example the PID histogram obtained from the correlation of a MSGC with the CsI7 in sector 13 is presented in Fig.~\ref{deegar} inset. Fragment charges are well identified. The charge identification thresholds are around 0.8~-~1 MeV/u. Fragments with Z$>$9 are not visible in Fig.~\ref{deegar}, due to statistics and kinematic limitations ($<\theta> =  47^\circ$). The resolution and the identification properties have been verified to be comparable to the ones obtained with the analog chain~\cite{garf}.

To quantitatively evaluate the results one can calculate the Figure-of-Merit (FoM) by fitting the PID peaks with Gaussian functions~\cite{fom}:

$$ FoM = \frac {(X_2 - X_1)} {(FW_1 + FW_2)} $$
where X$_1$ and X$_2$ are the Gaussian mean values and FW$_1$ and FW$_2$ correspond to their full width at half maximum. The FoM quantitatively defines and compares the quality of the discrimination. Values of FoM larger than 0.7 correspond to well separated Gaussians with a peak to valley ratio $\approx$~2 and less than $\approx$~5\% of contamination by the neighboring charges.  The charge (or the mass) can be still identified when the FoM is of the order of 0.4$\div$0.5 with larger contamination probabilities.
For the PID in the inset of Fig.~\ref{deegar} one gets values of the FoM ranging from $\approx$ 1 to 2.

\subsubsection{Drift and rise time of drift chamber signals} \label{psamic}

\begin{figure}[htb]
\centering
\resizebox{1.0\columnwidth}{!}{\includegraphics{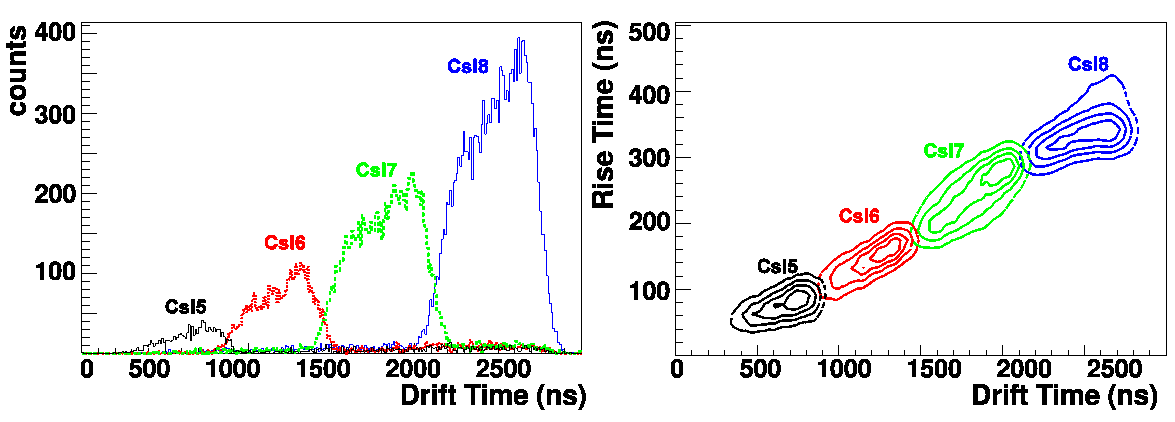}}
\caption {\it (Color online) Left panel: Drift times in coincidence with CsI 5 to 8 for one of the outer microstrips. Data refer to
fragments (Z$\ge$3).
 Different colors in the
time spectra correspond to the four CsI(Tl) of Fig.~\ref{driftfig}.
 Right panel: Rise time vs Drift Time correlation.
  Loci associated to different CsI(Tl) scintillators are well separated.}\label{ddig}
\end{figure}

As previously explained, the polar angle information can be extracted using the electron drift time.  Light charged particles give small signals, due to the low energy loss in the gas. If the signals are comparable to the noise of the electronic chain, it is more difficult to obtain a definite angular information. The fast amplification stadium of the N568 amplifiers, together with the C208 CFD threshold procedure determination, made it possible to obtain sufficiently precise drift times only for fragments with Z$\ge$3 in the analog case. The same result was easily obtained in the case digital electronics as shown in Fig.~\ref{ddig}, where the data for the drift time of fragments with Z$\ge$3 emitted in the reaction $^{32}$S + $^{40}$Ca at 17 MeV/u beam energy are displayed: the events are sorted as a function of each CsI(Tl) scintillator in the sector hit by the impinging particles and the well separated contributions in the spectrum are represented with different colors. The shortest times correspond to CsI5 of Fig.~\ref{driftfig}, whereas the largest to
CsI8.
Moreover with digital electronics, a clean drift time information can also be extracted for $\alpha-$particles, due to a better signal to noise ratio of the digitizing electronics.
\vspace{0.3cm}

\begin{figure}[htb]
\centering
\resizebox{1.0\columnwidth}{!}{\includegraphics{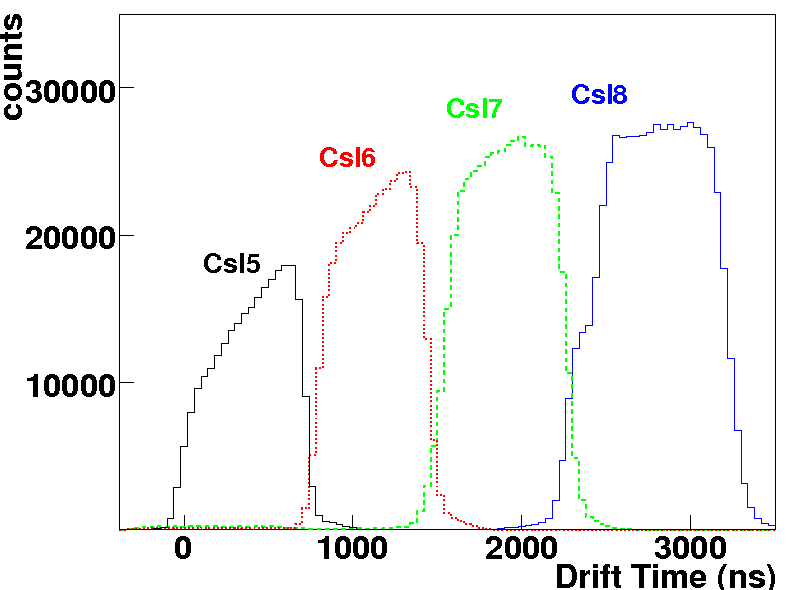} \includegraphics{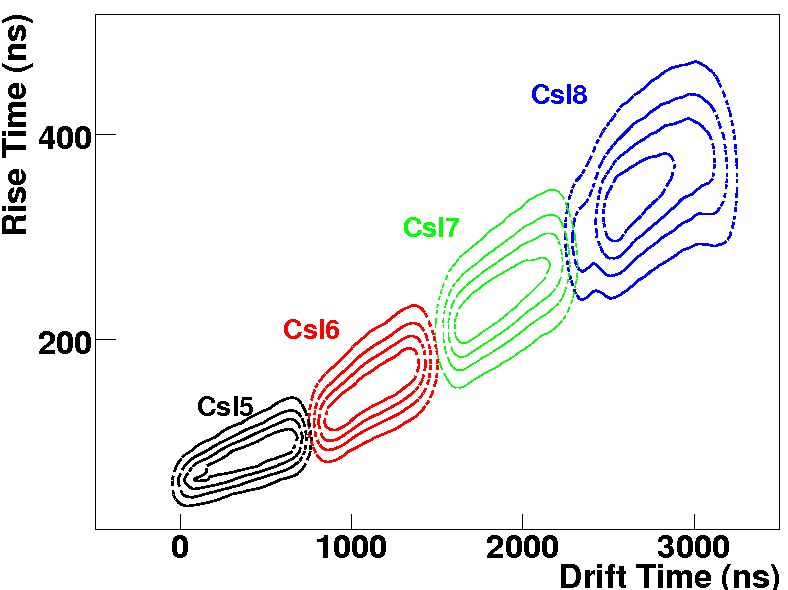}}
\caption {\it (Color online) Left panel: Drift times in coincidence with CsI 5 to 8 of one sector. Data refer to Z=2 fragments (identified using the
  $\Delta$E-E technique).
   Right panel: Rise time vs Drift Time correlation.   Loci associated to different CsI(Tl) scintillators are well separated.
  }\label{driftalpha}
\centering
\end{figure}

In Fig.~\ref{driftalpha}~(left panel) $\alpha-$particle data are shown, which have been collected by one of the outer microstrips for one GARFIELD sector and selected through the $\Delta$E-E identification technique. The energy of the $\alpha$-particles ranges from 8 to 130 MeV.

In the right panels of Figs.~\ref{ddig}~and~\ref{driftalpha} the correlation between rise time and drift time is presented. The two quantities are clearly correlated, even though the drift time shows up a major sensitivity to the polar angle, as expected. From the finite width of the borders of each CsI(Tl) drift time spectrum in Figs.~\ref{ddig}~and~\ref{driftalpha} we can estimate the angular resolution. It is found to be in the 1$\div$3 degree range in agreement with previous tests~\cite{garf}.

Obtaining the angular information also for low ionizing alpha particles is one of the advantages of using the present digital electronics in GARFIELD.

\subsubsection{Pulse Shape Analysis of CsI(Tl) signals} \label{psacsgar}

\begin{figure}[!h]
\centering
\resizebox{0.4\textwidth}{!}{\includegraphics{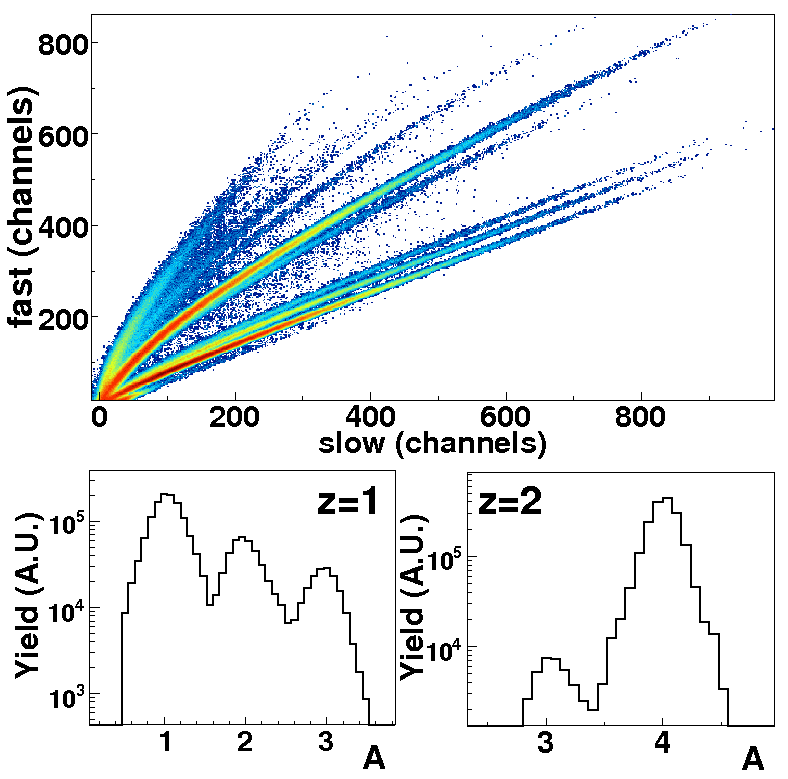} }
\centering
\caption {\it (Color online) Top panel: Fast-Slow bi-dimensional plots of one of the GARFIELD scintillators. One can recognize from the bottom the lines for p,d,t,$^3$He,$\alpha$, the line corresponding to the two $\alpha$'s coming from $^8$Be and the $^6$Li,$^7$Li lines. Bottom panels:
mass distributions for Z = 1 and Z = 2 isotopes.}
\label{fsmasg}
\end{figure}

It is well known that CsI(Tl) allows for particle identification through Pulse Shape Analysis (PSA).  The Fast-Slow technique was applied for GARFIELD according to the procedure described in Sect.~\ref{algcsig}~\cite{mor}. In Fig.~\ref{fsmasg} (top panel) Fast-Slow bi-dimensional plots for one of the CsI(Tl) detectors are shown: the Z-values can be identified up to Z~=~4, whereas the masses are clearly seen for Z~=~1 and Z~=~2; starting from the lowest line one can identify the $\gamma$-rays, then the three lines related to Z~=~1~(p,d,t), the two lines for Z~=~2~($^3$He, $\alpha$) and the $\alpha$ double hits which are associated mainly to the $^8$Be$_{g.s.}$ decay. Above this line the masses can hardly be identified for the two isotopes of Z~=~3, while the two remaining lines can be attributed to Z~=~4 and Z~$>$~4. In the bottom panel of Fig.~\ref{fsmasg} the mass distributions obtained following the procedure described in Ref.~\cite{mor} are shown. The energy range of $\alpha$-particles ranges from 8 to 130 MeV.

The FoM obtained for Z=1 are 0.98 and 0.96 for p-d and d-t, respectively. For Z=2 one gets FoM=1.29 for
$^3$He-$\alpha$ discrimination.
An automatic procedure has been studied to quickly identify the isotopes and to speed up the calibration phase~\cite{mor}.
Dedicated measurements have been performed for a precise energy calibration of the CsI(Tl) crystals: in this way the Light Output response function with respect to the energy and charge of the reaction products has been determined.

\section{Ring Counter}\label{rco}

The RCo is a three stage annular detector array with a truncated cone shape covering the $\theta =5^\circ \div 17^\circ$ angular range. The first stage is an ionization chamber (IC), followed by silicon strip detectors (Si) and CsI(Tl) scintillators.
The ion chamber is divided into eight azimuthal sectors (each one covering $\Delta \phi = 45^\circ$) sharing an unique gas volume. Eight separate pie shaped silicon detectors are mounted on an aluminum holder behind every IC sector: each of them is segmented into eight independent annular strips.
The third stage of the RCo has been renewed in order to increase the granularity: the two old CsI(Tl) crystals positioned after each silicon pad have been replaced by six smaller size 4.5 cm thick CsI(Tl) crystals (for a total number of 48 devices read by photodiodes). The geometrical shapes of the new CsI(Tl) crystals precisely fit the angular cone subtended by the IC.

A remotely controlled collimator system has been implemented in order to apply different screens in front of the RCo under vacuum and therefore while running the experiment. Different screens are used for focusing, shielding or collimation purposes. The RCo is mounted on a sliding plate and can be remotely moved back and forth in order to replace the front collimator. The RCo operational position is inside the forward cone of the GARFIELD drift chamber at 177 mm from the target as shown in Fig.~\ref{tot}.
The pre-amplifiers are mounted on the same sliding plate of the RCo close to the detectors, in order to reduce their input capacitance. They are thermally connected to shielded metallic boxes to be refrigerated through a water cooling system.

\subsection{The ionization chamber}\label{ic}

The IC is 6 cm long with three thin electrodes (1.5 $\mu$m aluminized mylar): an intermediate anode and two grounded entrance and exit cathodes. At present the detector is the same of the first version~\cite{rco} but novel signal processing is performed using the same digital electronics described for GARFIELD.
At the operating pressure of $\sim 50$ mbar, a resolution of about 6\% is obtained for the energy lost by elastically scattered 17 MeV/u $^{32}$S ions (their average energy loss in IC being $\approx$ 6 MeV).

\subsection{The silicon detectors}\label{sil}

The performance of the silicon detectors has been substantially improved. On one hand the new digitizing electronics has added pulse shape identification capabilities to the apparatus (e.g. by measuring both the amplitude and the rise-time of each pre-amplifier signal)~\cite{psa}. On the other hand, the study performed within the FAZIA framework ~\cite{fazia_web} demonstrated the better performances achievable using silicon detectors with good resistivity uniformity (less than few percent)~\cite{bard}. For this reason new nTD silicon detectors~\cite{can} (which are known to feature a high controlled doping homogeneity) have been mounted. The major advantage of a better homogeneity is the reduction of the pulse shape dependence on the particle interaction zone in the detector~\cite{luigi_fazia}. Furthermore, at present the silicon detectors have been reverse mounted: the pulse shape discrimination capability is, in fact, enhanced by forcing the particle to enter through the low field side~\cite{carboni_fazia,nicol}.

Each silicon pad covers one sector and the rear electrode (junction side) is segmented into eight strips.
The polar angle interval covered by each strip is quoted in Table~I. Values in the Table refer to silicon detectors of the RCo in the measuring position, namely at 279 mm from the target.

\begin{figure}[htb]
\centering
\resizebox{0.9\columnwidth}{!}{
\includegraphics{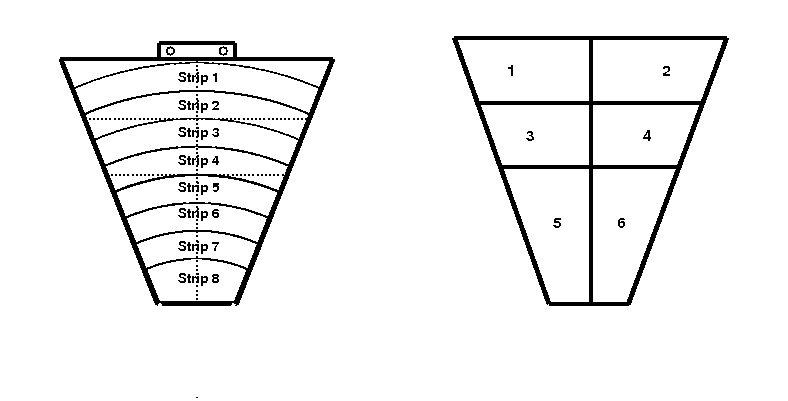}
}
\centering
\caption{\it Left panel: scheme of the strips of the silicon detector. The dotted lines correspond approximately to the different CsI(Tl). Right panel: CsI(Tl) scintillator limits for one of the eight sectors of the RCo.}
\label{rcosicsi}
\end{figure}

\begin{center}
\begin{tabular}{ccccc}
\multispan {5} {\bf Table I - Radii and polar strip angles}
\\
  \hline
  strip & int. radius  & ext. radius  & min. angle & max angle \\
   & (mm)  &  (mm) & (deg)& (deg) \\
   \hline
1        &77.9      &85.0    & 15.6  & 17.0   \\
2        &70.8      &77.8    & 14.2  & 15.6   \\
3        &63.7      &70.7    & 12.9  & 14.2   \\
4        &56.6      &63.6    & 11.5  & 12.8   \\
5        &49.4      &56.4    & 10.1  & 11.4  \\
6        &42.3      &49.3    & 8.6  & 10.0  \\
7        &35.2      &42.2    & 7.2  & 8.6  \\
8        &26.2      &35.1    & 5.4  & 7.2   \\
  \hline \label{radii}
  \end{tabular}
\end{center}
 A simple sketch of the silicon is shown in the left part of Fig.~\ref{rcosicsi}, in which one can also see the dotted lines corresponding approximately to the borders between different CsI(Tl) crystals. Strip 8 is different from the others because the silicon is cut with a straight line. This does not affect the shape of the active strips since in front of the RCo we have a circular collimator which covers part of the inner strip (sometimes, considering the grazing angle, the inner strip is completely or partially screened.)

Each strip covers a solid angle of 2~$\div$~6~msr. Each pad is completed by a guard ring, mounted all around the strips and properly biased, which is needed to minimize the field distortion effects in the inter-strip regions (220 $\mu$m wide). In this way charge split and cross-talk are drastically reduced. The bulk resistivity of the detector is of the order of 3400 $\Omega\cdot$cm, the full depletion voltage is 100 V with recommended bias of 120 V.
The reverse current of each strip is about 30 nA. The silicon detectors feature relatively thin window thicknesses (50~nm junction side, 350~nm Ohmic side) in order to get low thresholds even in the reverse mounting configuration. The thickness of the silicon detectors is around 300 $\mu$m, and therefore the punching through energies are about 6 MeV/u for protons and $\alpha$-particles and 7$\div$11 MeV/u for light fragments. A resolution of the order of 0.3\% both for 5 MeV $\alpha$ particles and for elastically scattered 17 MeV/u $^{32}$S (about 500 MeV residual energy) has been obtained.
The preamplifiers used for the silicon strips have a gain of 5~mV/MeV and are fed into the same digital read-out cards used for the other channels.

\subsection{The CsI(Tl) scintillators}\label{csi}

\begin{figure}[htb]
\centering
\resizebox{0.9\columnwidth}{!}{
\includegraphics{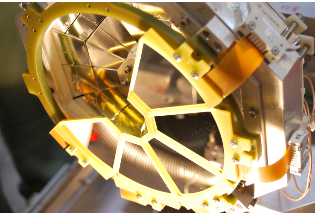}
}
\centering
\caption{\it (Color online) Picture of the new RCo geometrical configuration during its mounting. Five sectors of Silicon Detectors are visible in  the front plane, behind them (upper-left side) one can see the CsI(Tl) wrapped crystals.}
\label{csi_config}
\end{figure}

The granularity of RCo has been increased bringing the number of CsI(Tl) scintillators  from 16 to 48. For each sector the {\it 4 Si inner strip} polar angular region is now azimuthally divided in two parts: in place of the previous  single crystal ($\Delta \phi$ = 45$^\circ$), two CsI(Tl) scintillators are located ($\Delta \phi$ = 22.5$^\circ$ each). Moreover the {\it 4 outer strip} region is now covered by four CsI(Tl) crystals, each covering half of the polar and half of the azimuthal angular range defined by the strips. The smaller dimension of the CsI(Tl) allows for better doping (and therefore scintillation) uniformity and better light collection of the single detector (better photodiode area to crystal dimension matching). Having scintillators of similar size also favors response uniformity among the various detectors. The right panel of Fig.~\ref{rcosicsi} shows a front cross section of the position of the six CsI(Tl) of each sector. A picture of the upgraded RCo during its mounting is shown in Fig.~\ref{csi_config}. In the region where the silicon detectors were not yet mounted the upgraded CsI(Tl) configuration can be seen.

As a consequence of studies on the resolution and Light Output performances of CsI(Tl) scintillators as a function of the doping content and homogeneity~\cite{wag,csi2}, the present CsI(Tl) crystals have been chosen with a Tl doping in the range of 1500$\div$2000 ppm. The 48
crystals, 45 mm thick, have been wrapped with micro-porous polyvinylidene fluoride membrane~\cite{mil} and Teflon tape. They have been optically coupled with Hamamatsu photodiodes (S2744-08, 10 x 20 mm$^2$), 500 $\mu$m thick, dark current $\leq$ 15 nA).

The preamplifiers have the same characteristics as the ones described for GARFIELD.

\subsection{Algorithms for IC, Si and CsI(Tl) of RCo}\label{rco_csi_algo}

In order to extract the energy information from the IC signals, the DSP performs the same semi-Gaussian shaping algorithm as employed for the GARFIELD chamber~(see Sec.~\ref{algdrift}). Even in this case the shaper is applied to a decimated version (128~ns sampling period instead of 8~ns) of the IC pre-amplifier signal~\cite{dig1}.

A {\it time of occurrence} information is also extracted from IC signals, using a dCFD with a 25\% threshold. This information is used to recognize spurious signals, i.e. due to {\it other bunch} interactions and therefore not in coincidence, to be discarded in the analysis.

The algorithms employed in the analysis of silicon detector signals are the same as those applied for GARFIELD micro-strips (see Sec.~\ref{algdrift}). Energy, time of occurrence and rise time information are extracted. The main parameters, in this case, are the amplitude (i.e. energy) and rise time information: their correlation allows for fragment identification via Pulse Shape Analysis.

The algorithms employed in the analysis of the CsI(Tl) scintillators are the same used for the GARFIELD scintillators (see Sec.~\ref{algcsig}).

\subsection{Performances of the detectors of the RCo} \label{perf}

Two complementary methods can be used to identify the reaction products.
\begin{enumerate} \item the $\Delta$E-E technique. Considering the RCo three fold stage one can extract information on:
\begin{itemize}
\item particles stopped in the silicon detector: from the energy lost in the ionization chamber $vs$ the residual energy in a silicon strip;
\item particles stopped in the CsI(Tl) scintillators: from the energy lost in a silicon strip $vs$ the  residual energy in the scintillator.
\end{itemize}
\item the Pulse Shape Analysis which gives information on:
\begin{itemize}
\item particles stopped in the silicon detector through energy - risetime correlations;
\item particles stopped in CsI(Tl) through fast-slow correlations.
\end{itemize}
\end{enumerate}

\subsubsection{$\Delta$E - E from IC - Si} \label{deerco1}

\begin{figure}[htb]
\centering
\resizebox{0.48\textwidth}{!}{\includegraphics{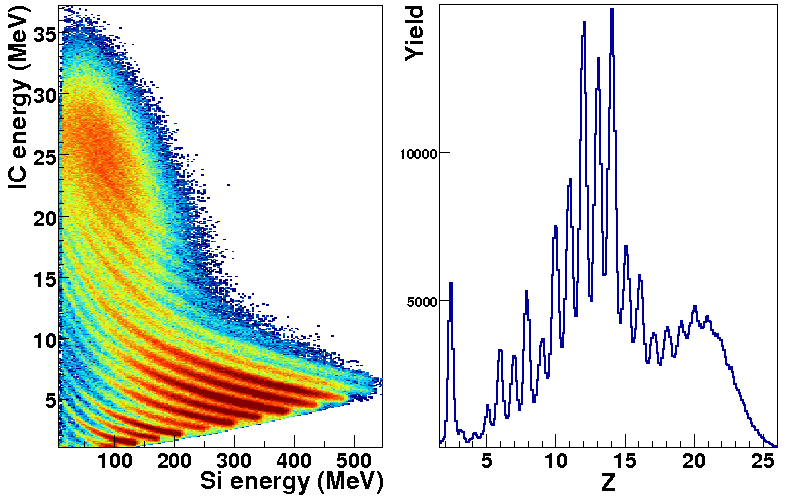}
}
\centering
\caption  {\it (Color online) Left panel: $\Delta$E - E correlation ``IC vs silicon'' for the reaction $^{32}$S + $^{40}$Ca at 17 MeV/u at $\theta \approx$ 8$^o$. The line which extends to the largest silicon energy corresponds to $^{32}$S ions. Right panel: PID for the same data. The Z~=~2 distribution is slightly shifted from the right value since the procedure is optimized for greater Z values.
}
\label{icsi}
\end{figure}

 A typical $\Delta$E~-~E scatter plot for a IC sector signal ($\Delta$E) and a silicon strip signal (E) correlation is shown in Fig.~\ref{icsi} (left panel) for the fragments produced in the reaction $^{32}$S + $^{40}$Ca at 17 MeV/u. Charged particles and fragments are clearly visible, together with low energy heavy residues formed through fusion reactions.

The different nuclear species can be identified in charge and their kinetic energy can be obtained, after proper calibration. Especially for the low energy region of the heavier fragment, the "in gas" energy loss corrections, estimated according to Refs.~\cite{texas,latt} have to be  considered. The high quality of the detectors and electronics can be demonstrated by the wide range of charge identified ions: as shown in Fig.~\ref{icsi} the ridges are rather separated up to Scandium, corresponding to the evaporation residue region and down to 0.8$\div$1.0 MeV/u in energy.
In the right panel of Fig.~\ref{icsi} the PID charge identification function is shown. The FoM for increasing charge of the identified fragments to the beam charge varies from 0.8 to 0.5, whereas in the residue region (Z around 20) the FoM becomes of the order of 0.4.

\subsubsection{$\Delta$E - E plots from Si - CsI(Tl)} \label{deerco2}

\begin{figure}[!h]
\centering
\resizebox{0.4\textwidth}{!}{\includegraphics{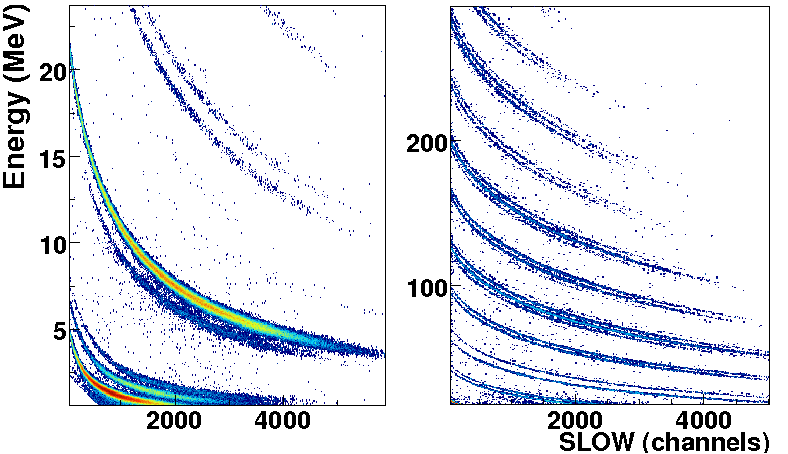} }
\centering
\caption
{\it (Color online) $\Delta$E-E correlations obtained for a representative pair of a silicon strip and a CsI(Tl).
The left panel shows the
resolution for Z = 1$\div$3, whereas the right panel focuses on ions
Z $>$ 3. The two well separated isotopic lines of $^7$Be and $^9$Be are easily
recognized in the right panel, starting at a $\Delta$E energy of $\simeq$50 MeV. The maximum energy for $\alpha$-particles is around 140 MeV, for Carbon 270 MeV and for Neon 430 MeV.}\label{sicsi}
\end{figure}
Thanks to the excellent energy resolution of the Silicon detectors, the Si-CsI(Tl) $\Delta$E-E correlation allows to identify the reaction  products both in Z and A. In particular it is evident from Fig.~\ref{sicsi} that the mass separation is certainly achievable up to Z~=~11$\div$12. Even though the available statistics in the studied reaction does not allow to determine the upper Z value of mass separation, a quantitative evaluation of the quality of the detector resolution can be inferred from the FoM.

\begin{figure}[!h]
\resizebox{0.46\textwidth}{!}{%
\centering
  \includegraphics{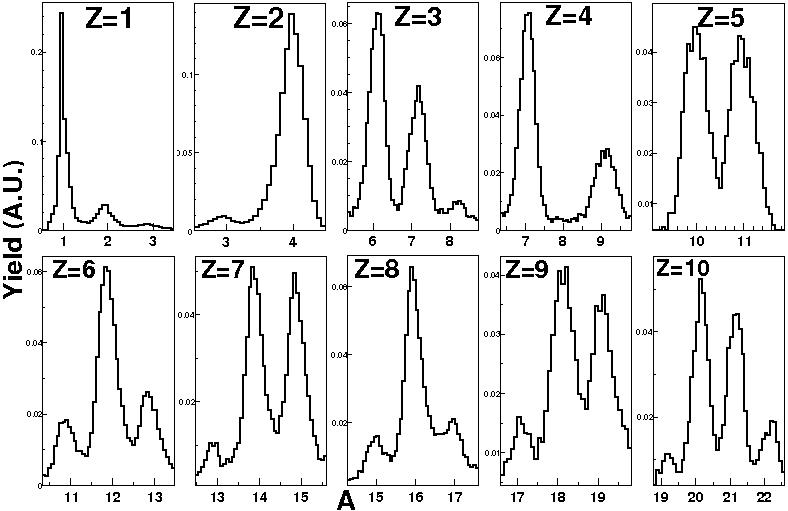}}
\caption
{\it Mass identification with the procedure described in Ref.~\cite{len}.}\label{mass}
\centering
\end{figure}

In Fig.~\ref{mass} the results of mass identification using the procedure described in Ref.~\cite{len} is shown. The obtained FoM ranges from 1.1 at Z = 1 down to 0.77 at Z = 10, showing that Z=10 is still not at the upper limit for good mass separation.

\subsubsection{PSA in Silicon detectors}\label{psasi}

As already mentioned, the reverse mounting of the silicon detectors allows for a better exploitation of PSA. Moreover the use of the
fast digitizing electronics allows for extracting several parameters from the signal waveform, a few of them well suited for PSA.

\begin{figure}[h]
\centering
\resizebox{0.4\textwidth}{!}{
\includegraphics{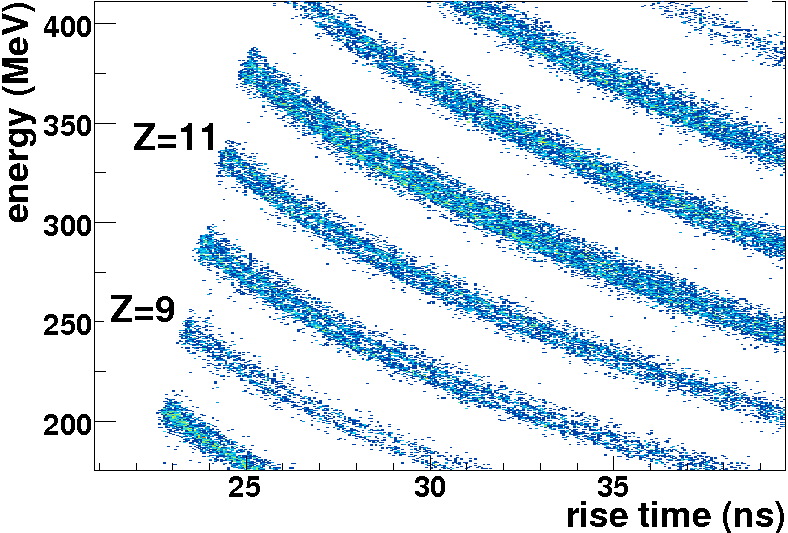}}
\caption
{\it (Color online) Upper panel: PSA ``Charge vs Risetime'' for a silicon detector strip of the RCo. Bottom panel: Mass identification for a restricted Charge vs Risetime region }\label{psaall}
\centering
\end{figure}

In Fig.~\ref{psaall} the energy $vs.$ rise-time signals have been plotted ~\cite{psa,carboni_fazia}.
Data refer to the results of a typical detector strip. A good charge separation has been achieved for all the products. The lower energy threshold for charge identification is $\approx~$2.5$\div$5 MeV/u, which corresponds approximately to the energy needed by the fragment to travel along the first 30 $\mu$m of the detector (see. Ref.~\cite{luigi_fazia,nicol}).

 Zooming the energy-risetime plot it appears that some marginal mass resolution is achievable in some regions of the correlation plot (see Fig~\ref{psaall} bottom panel). Efforts are in progress in order to improve the resolution in view of future experiments. In that respect the use of preamplifiers of higher sensitivity is promising~\cite{luigi_fazia}. In fact
the ion separation capability of silicon via PSA strongly depends on the risetime determination and this latter is influenced by the noise especially for small signal amplitudes. Increasing the preamplifier sensitivity gives better signal to noise ratio, taking into
account the sampling ADC internally generated noise. As a consequence, a
better precision in
risetime measurements can be achieved. Fig~\ref{psaall}  shows  the quality of the separation for the same RCo detector when using a larger sensitivity (45mV/MeV@Si): the improved behavior clearly appears,  even if some saturation effect is present.

\begin{figure}[h]
\centering
\resizebox{0.4\textwidth}{!}{
\includegraphics{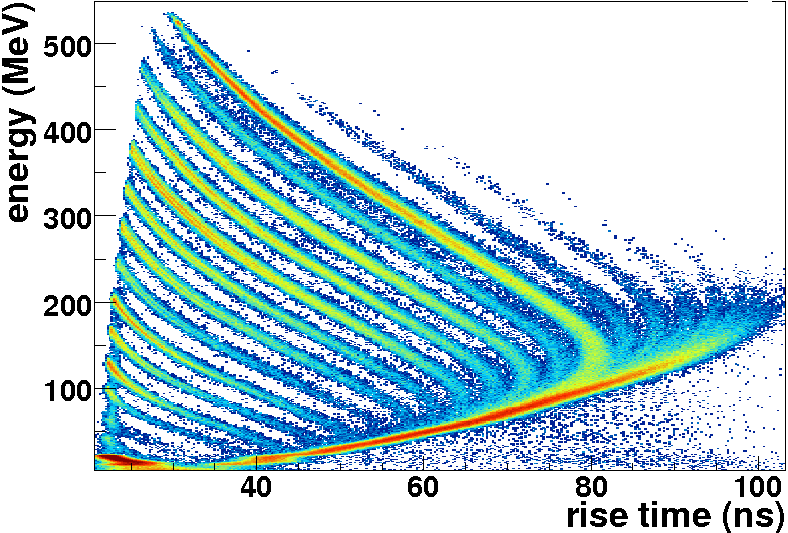}}
\centering
\end{figure}
\vspace{0.5 cm}

\begin{figure}[htb]
\centering
\resizebox{0.4\textwidth}{!}{%
  \includegraphics{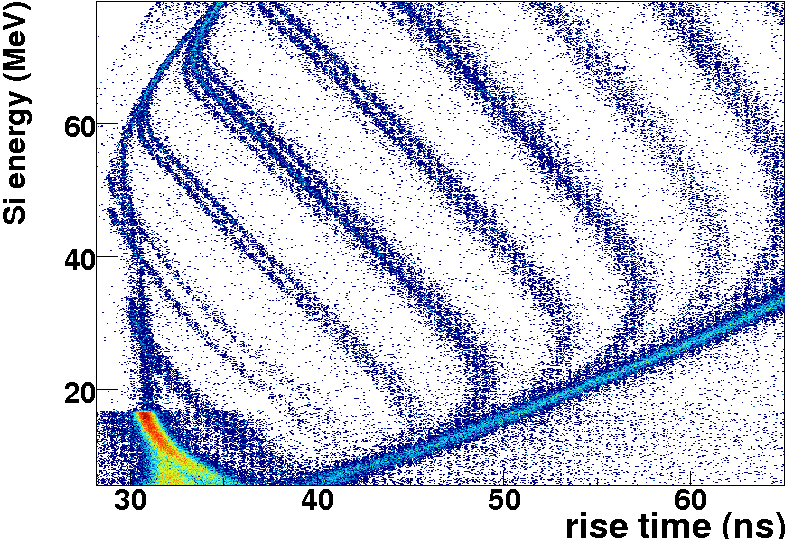}}
\caption
{\it (Color online) PSA with a high gain pre-amplifier.}\label{psamass}
\centering
\end{figure}

\subsubsection{Pulse Shape Analysis in CsI(Tl) detectors}
\label{psacsi}

\begin{figure}[!h]
\centering
\resizebox{0.95
\columnwidth}{!}{\includegraphics{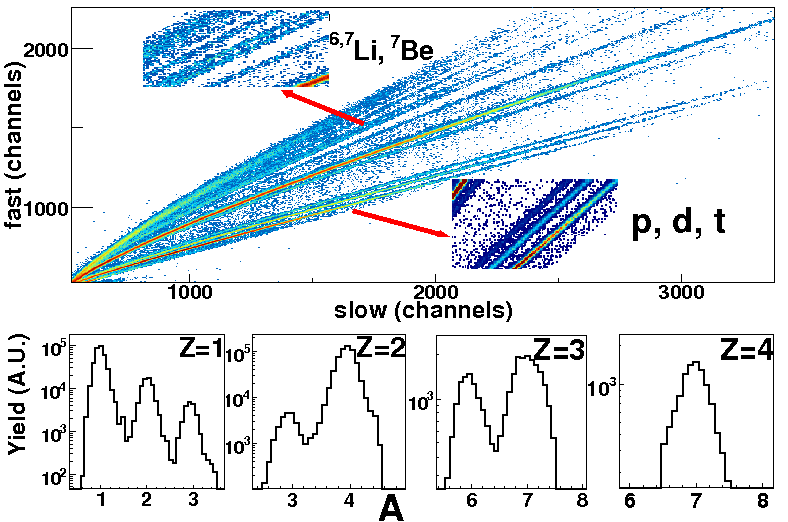}}
\centering
\caption
{\it (Color online) Upper panel: Fast slow correlation for one of the
CsI(Tl) of the RCo.
Lower panels: Results obtained for mass identification of the same CsI(Tl) with the
procedure described in Ref.~\cite{mor}}\label{csi6}
\end{figure}

As for the GARFIELD scintillators the PSA method (see Sect.~\ref{psacsgar}) is applied to the RCo CsI(Tl) signals. Typical results are shown in Fig.~\ref{csi6} (top panel). In the insets a zoom for Z=1 and Z=3,4 shows the achieved mass identification. The $\alpha$-particle energy ranges from 25 to 130 MeV.

In the lower panels of Fig.~\ref{csi6} the mass identification obtained for Z = 1 up to Z = 4 in the whole energy range is shown (for Z = 4 only the isotope $^7$Be is well separated and identified).
The calculated FoM for masses of Z = 1 and Z = 2 particles is larger than 1.5 and for Z = 3 it is 0.82.

\section{Results}

\subsection{Summary of the GARFIELD upgrading} \label{sumg}

The results shown in Sect.\ref{perfgar} indicate that the GARFIELD detector allows for:
\begin{description}
\item{i.} energy determination with accuracy of the order of few percent;
\item{ii.} $\Delta$E-E charge identification with an energy threshold of 0.8$\div$1 MeV/u;
\item{iii.} low threshold ($\simeq$~2$\div$4 MeV/u) mass identification of light isotopes (up to Z = 3) in the  whole GARFIELD angular range through the pulse shape analysis of the CsI(Tl) detectors;
\item{iv.} polar angle resolution of $\Delta \Theta = 1^\circ \div3^\circ$ and azimuthal angle determination of   $\Delta \phi = 7.5^\circ$~\cite{garf}.
\end{description}

The new digitizing electronics allows for the extraction of all the relevant information on energy, drift time and  pulse shape discrimination: part of this information was not available with the standard electronic modules previously employed. The digital FEE fully replaced analog shapers, CFDs, peak-sensing ADCs, programmable delays and standard TDCs and it removed the need for duplicated electronic chains (low and high gain). This constitutes an undoubtable advantage in the experimental setup handling.

\subsection{Summary of the RCo upgrading} \label{sumRCo}
The results shown in Sect.~\ref{perf} indicate that the RCo detector allows for:
\begin{description}
\item{i.} energy determination with accuracy of the order of few percent;
\item{ii.} charge identification of particles and fragments through $\Delta$E-E (identification threshold = 0.8$\div$1 MeV/u) and energy {\it vs.} risetime correlation. The two methods have different features: the identification thresholds are high\-er for the PSA with respect to the $\Delta$E~-~E; on the contrary the FoM obtained by the PSA are better than those obtained by $\Delta$E~-~E due to the resolution of the gas $\Delta$E. In addition if two particles/fragments enter in the same IC sector, but in different silicon strips, only the PSA can discriminate between them.  The obtained results with PSA will make it possible to perform specific experiments at very low IC pressure, or even to remove it, depending on  the charge-energy distributions of the ions to be detected.
\item{iii.} mass identification of fragments (up to at least Z = 11) from the Si-CsI(Tl) $\Delta E$-E correlation and Pulse Shape - risetime correlation;
\item{iv.} mass identification of light isotopes (up to Z = 3) exploiting the Pulse Shape Analysis of CsI(Tl) crystals, with thresholds of the order of $\simeq$~6 MeV/u, due to the preceding 300 $\mu$m silicon detector.
\end{description}

\subsection{Results of the coupled apparatuses}

In order to show which kind of physics the coupled apparatuses are able to perform, we point to the high quality of the charged particle identification and of the energy determination. Indeed if one calculates the correlation functions between couples of identified isotopes one can infer the contribution of excited levels decaying to a given pair of isotopes. In a recent paper an experiment performed with GARFIELD and RCo apparatuses in their first version~\cite{stag} has demonstrated that the position of the levels of excited parent nuclei can be well recognized by correlation functions. The further improved granularity and particle identification obtained with the present version of the apparatuses will allow to perform this analysis with a reduced distortion and smaller width of the correlation peaks.

\begin{figure}[!h]
\centering
\resizebox{0.32\textwidth}{!}{\includegraphics{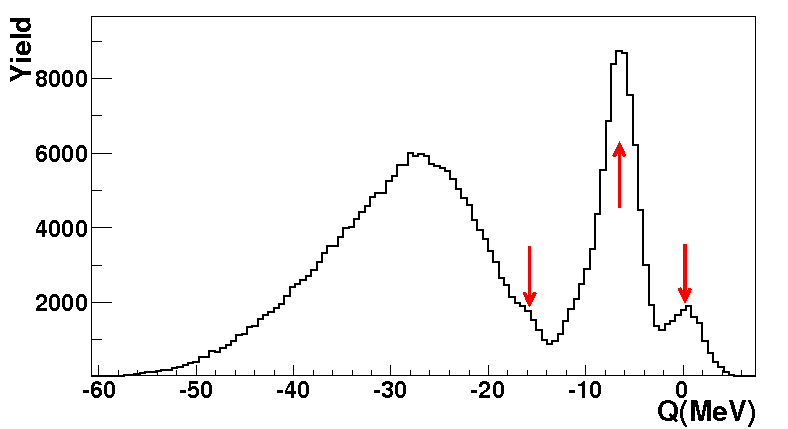}}
\caption
{\it (Color online) Q-values of reactions $^{12}C$($^{12}C$,$\alpha \alpha$)$^{A}O$. The arrows correspond to $\alpha$-decay chains, starting from the $^{24}$Mg$^*$ compound nucleus and leaving an $^{16}$O residue either in its ground state or in one of its excited bound states at $E^*$= 6.05, 6.13, 6.92 and 7.12 MeV. The leftmost arrow corresponds to the upper limit ($Q=-15.78$) of the opening of the 4-body channel going to $^{12}C$($^{12}C$,$\alpha \alpha$n)$^{15}O$. For more details see text.}\label{qv}
\end{figure}

The overall quality of the
detector response (namely mass, energy and position
determination) is tested by an example of Q-value
estimation from a 3-body event reconstruction~\cite{latt}.
Indeed in a recent experiment~\cite{cc} performed with GARFIELD + RCo, the Q-values of a specific reaction, i.e. $^{12}C$~+~$^{12}C$ going to $^{A}O$~+~$\alpha$~+~$\alpha$, show values compatible with the calculated ones from the mass difference, as it is clearly shown in Fig.~\ref{qv}, where one can see three different peaks. The rightmost peak corresponds to the reaction $^{12}C$($^{12}C$,$\alpha \alpha$)$^{16}O$, leaving the $^{16}$O residue in its ground state ($Q= -0.11$ MeV). The central peak corresponds to the reaction $^{12}C$($^{12}C$,$\alpha \alpha$)$^{16}O^*$, with the $^{16}$O residue in one of its excited bound states at $E^*$= 6.05, 6.13, 6.92 and 7.12 MeV, which are so close in energy that cannot be resolved in the experiment and the measured average energy expense is $Q\approx-6.55$ MeV. (We note that
this estimation largely bases on data from the CsI detectors
and therefore cannot be compared with specific high-resolution
spectrometers). The leftmost arrow corresponds to the opening  ($Q=-15.78$) of the reaction $^{12}C$($^{12}C$,$\alpha \alpha$n)$^{15}O$, where neutrons, not detected in this experiment, have a broad energy distribution.

\section{Conclusions}\label{concl}

The improved GARFIELD + RCo apparatus has been described.
The array, designed to study nuclear reactions in the 5$\div$20 MeV/u energy range at LNL, will be employed also in future experiments with the radioactive SPES beams. The main properties of the setup are the large acceptance, the good energy resolution, the low energy thresholds and the high granularity especially in the forward part (RCo) which allow for correlation function studies. Moreover the GARFIELD+RCo complex is the first european 4$\pi$ charged-particle detector array fully provided with digital electronics. The new digitizing boards allowed for a sensible reduction of the needed front-end channels, keeping or even improving  the performances obtained with the analog electronics.\\
Different techniques are used to identify particles, such as $\Delta$E-E correlations and Pulse Shape Analysis in scintillators and silicon detectors: these complementary techniques allow for extensive charge and, for the lightest products, mass identification in a large energy and angular range.

This improved apparatus well matches the needs of advanced studies on nuclear reactions, especially where the complete event reconstruction is required~\cite{stag,cc}. For example the studies of pre-equilibrium LCP emission~\cite{krav} can now be extended to Z = 1 and Z = 2 isotopes, allowing deeper comparisons with models. A better reconstruction of warm fragments~\cite{stag} via correlation functions will be possible with the extended mass identification and the increased granularity.

\vspace{0.5cm} {\bf Acknowledgments}

The authors are indebted to R. Cavaletti, L. Costa, M. Ottanelli, A. Paolucci, S. Serra, G. Tobia and A. Zucchini for their skillful assistance.
Thanks are due to the accelerator staff of Legnaro Laboratories for having provided good quality beams. This work was partly supported  by grants of Alma Mater Studiorum (Bologna University), ENSAR Grant Agreement n. 262010 - Combination of Collaborative Project \& Coordination and Support Actions and Fondazione Comunicazione e Cultura (CEI).

\end{document}